\documentstyle[psfig]{mn2e}
\newcommand{\etal}{{et al.~}}
\newcommand{\lta}{\la}

\newcommand{\kmsmpc}{\>{\rm km}\,{\rm s}^{-1}\,{\rm Mpc}^{-1}}
\newcommand{\kms}{\>{\rm km}\,{\rm s}^{-1}}

\newcommand{\kpc}{\>{\rm kpc}}
\newcommand{\Msun}{\>{\rm M_{\odot}}}

\newcommand{\beq}{\begin{equation}}
\newcommand{\eeq}{\end{equation}}

\newcommand{\msunh}{\>h^{-1}\rm M_\odot}

\newcommand{\apj}{ApJ}

\newcommand{\aj}{AJ}
\newcommand{\mnras}{MNRAS}
\newcommand{\aap}{A\&A}

\newdimen\hssize
\newdimen\hdsize 
\begin{document}
            

\title[The Alignment between Satellites and their Central Galaxy]
      {The Alignment between the Distribution of Satellites and
       the Orientation of their Central Galaxy}
\author[X. Yang et al.]
       {\parbox[t]{\textwidth}{
        Xiaohu Yang$^{1,2,7}$ \thanks{E-mail: xhyang@shao.ac.cn}, 
        Frank C. van den Bosch$^{3}$, H.J.~Mo$^2$,
        Shude Mao$^4$, Xi Kang$^5$,\\Simone M. Weinmann$^6$, 
        Yicheng Guo${^2}$, Y.P. Jing$^{1,7}$}\\
        \vspace*{3pt} \\
       $^1$Shanghai Astronomical Observatory; the Partner Group of MPA,
           Nandan Road 80,  Shanghai 200030, China\\
       $^2$Department of Astronomy, University of Massachusetts,
           Amherst MA 01003-9305, USA\\
       $^3$Max-Planck-Institute for Astronomy, K\"onigstuhl 17, D-69117
           Heidelberg, Germany\\
       $^4$University of Manchester, Jodrell Bank Observatory,
           Macclesfield, Cheshire SK11 9DL, UK\\
       $^5$Astrophysics Department, University of Oxford, Oxford OX1
           3RH, UK\\
       $^6$Institute for Theoretical Physics, University of Z\"urich,
           CH-8057, Z\"urich, Switzerland\\
       $^7$Joint Institute for Galaxy and Cosmology (JOINGC) of SHAO
           and USTC}


\date{}


\maketitle

\label{firstpage}


\begin{abstract}
  We use galaxy  groups selected from the Sloan  Digital Sky Survey to
  examine the alignment between  the orientation of the central galaxy
  (defined  as the  brightest group  member) and  the  distribution of
  satellite galaxies.  By construction,  we therefore only address the
  alignment on scales  smaller than the halo virial  radius. We find a
  highly significant  alignment of satellites  with the major  axis of
  their  central galaxy.  This  is in  qualitative agreement  with the
  recent  study  of Brainerd  (2005),  but  inconsistent with  several
  previous studies  who detected a preferential minor  axis alignment. 
  The  alignment  strength in  our  sample  is  strongest between  red
  central galaxies and red satellites.  On the contrary, the satellite
  distribution  in systems with  a blue  central galaxy  is consistent
  with isotropic. We also find that the alignment strength is stronger
  in  more massive  haloes and  at  smaller projected  radii from  the
  central  galaxy.   In addition,  there  is  a  weak indication  that
  fainter  (relative  to  the  central  galaxy)  satellites  are  more
  strongly  aligned. We  present a  detailed comparison  with previous
  studies,  and discuss the  implications of  our findings  for galaxy
  formation.
\end{abstract}




\section{Introduction}
\label{sec:intro}

\begin{table*}
\caption{The observational search for possible alignment between
central galaxies and their satellites.}
\begin{tabular}{lccccccc}
   \hline
Attempt & $N$ systems & central galaxy type & $r_p/\kpc$ & $|\Delta v|/\kms$ & $L_s/L_c$ & alignment & LG included  \\
 (1)    & (2)         & (3)                 & (4)        & (5)               & (6)       & (7)       & (8) \\
\hline\hline
Holmberg (1969)       & 218   & nearby spirals  & $\lta 50$        &  --              & --          & minor axis & YES\\
Zaritsky et al. (1997)& 115   & nearby spirals  & $\lta 200$       & $\lta 500$       & --          & NO         & -- \\
                      &       &                 & $[300~500]$      & $\lta 500$       & --          & minor axis & -- \\
Sales \& Lambas (2004)& 1276  & 2dFGRS BIG(*)   & $\lta 500$       & $\lta 160$       & $\lta 0.16$ & minor axis & -- \\
Kroupa et al. (2005)  & 11    & Milky-Way       & $\lta 250$       & --               & --          & minor axis & YES\\
Brainerd (2005)       & 3292  & SDSS-DR3 BIG    & $\lta 700$       & $\lta 1000$      & $\lta 0.16$ & major axis & -- \\
                      & 1575  &                 & $\lta 500$       & $\lta 1000$      & $\lta 0.25$ & major axis & -- \\
                      & 935   &                 & $\lta 500$       & $\lta 1000$      & $\lta 0.125$& major axis & -- \\
AB (**) (2005b)       & 4327  & SDSS-DR4 BIG    & $\lta 700$       & $\lta 500$       & $\lta 0.25$ & major axis & -- \\
This work (2006)      & 24728 &SDSS-DR2 GCG(***)&$\lta r_{\rm vir}$&$\lta v_{\rm vir}$& --          & major axis & -- \\
\hline
\end{tabular}
\medskip
\begin{minipage}{\hdsize}
  Column~(1)  indicates  the   attempt  ID.   Columns~(2)  (number  of
  central-satellite systems),~(3)  (the type of  central galaxy), ~(4)
  (the projected  centric-distance of the satellite  galaxy) ~(5) (the
  line-of-sight velocity difference  of the satellite-central system),
  and ~(6)  (the luminosity fraction of  the satellite-central system)
  indicate  the selection  criteria.  Column~(7)  lists  the alignment
  signal obtained from this  observation.  Column~(8) lists the status
  of the  Local Group.  (*)  BIG means Bright Isolated  Galaxies, (**)
  AB: Agustsson \& Brainerd (2005b)  and (***) GCG means Group Central
  Galaxy.
\end{minipage}
\end{table*}

During  the hierarchical  assembly of  dark matter  haloes, progenitor
haloes often survive accretion onto  a larger system, thus giving rise
to  a  population of  subhaloes.   If  the  baryonic material  in  the
progenitor  haloes  managed  to  cool  and  form  stars  before  being
accreted, the population  of subhaloes will give rise  to a population
of satellite galaxies.   Meanwhile, gas that cools onto  the center of
the parent halo gives rise to a so-called central galaxy.

Since  satellite galaxies  are typically  distributed over  the entire
dark matter  halo, they are ideally  suited as a  tracer population of
the potential well in which  they orbit.  Consequently, they have been
used  extensively  as  dynamical  tracers  of  the  dark  matter  mass
distribution surrounding  central galaxies.  In  addition to providing
accurate dynamical  masses of the  haloes (e.g., Zaritsky  \etal 1993,
1997a; McKay \etal 2002; Brainerd \& Specian 2003; van den Bosch \etal
2004),  the  radial trend  of  the  projected  velocity dispersion  of
satellite  galaxies can  also put  constraints on  the  radial density
distribution  of   the  dark  matter  (Prada   \etal  2003).   Similar
constraints can also come from the measurement of individual satellite
orbits, such as that of  the Large Magellanic Cloud or the Sagittarius
stream  in the Milky  Way halo  (e.g., Ibata  \etal 2001;  Helmi 2004;
Kallivayalil \etal 2006).

In  addition to these  kinematics, the  {\it spatial}  distribution of
satellite galaxies also holds  important information. If subhaloes are
a fair tracer of the dark  matter mass distribution, i.e., if they are
not spatially biased  in any way, the radial  and angular distribution
of satellite galaxies directly  reflects the projected distribution of
the dark matter.  If, on the  other hand, there is a spatial bias, the
satellite  distribution  holds important  clues  regarding the  actual
assembly history of the dark matter haloes.

Numerical  simulations  predict   that  the  spatial  distribution  of
subhaloes is  biased in two distinct  ways.  First of  all, the radial
distribution of  subhaloes is found to be  less centrally concentrated
than the dark matter, with  a pronounced deficit of subhaloes near the
center (Ghigna  \etal 1998, 2000;  Col\'in \etal 1999;  Springel \etal
2001; De  Lucia \etal 2004; Gao  \etal 2004; Diemand,  Moore \& Stadel
2004; Mao  \etal 2004;  Nagai \& Kravtsov  2005; Kang \etal  2005).  A
similar   prediction   has   also   been   obtained   with   detailed,
semi-analytical models  (Zentner \etal 2005a,  but see also  Taylor \&
Babul 2004).  Somewhat surprisingly, the observed spatial distribution
of  satellite galaxies,  especially in  clusters, appears  to  be more
centrally  concentrated  and  not  as  strongly  anti-biased  as  this
predicted distribution of subhaloes  (Carlberg, Yee \& Ellingson 1997;
van den  Marel \etal 2000; Lin,  Mohr \& Stanford 2004;  van den Bosch
\etal  2005;   Yang  \etal  2005c;  Chen  \etal   2005).   A  possible
explanation,  which  needs to  be  explored  in  detail, is  that  the
addition  of  baryons makes  the  subhaloes  more  resilient to  tidal
disruption.

In addition to this  radial anti-bias, numerical simulations have also
suggested an angular bias.  In particular, numerous studies have shown
that   dark  matter   haloes  in   dissipationless   simulations  have
anisotropic  distributions of  subhaloes that  are aligned  with their
major axis  (Knebe \etal 2004;  Zentner \etal 2005b; Wang  \etal 2005;
Libeskind  \etal 2005).   This anisotropy  mainly owes  to  a prefered
direction of  satellite accretion along large  scale filaments (Tormen
1997;  Vitvitska  \etal 2002;  Knebe  \etal  2004;  Aubert, Pichon  \&
Colombi  2004;  Zentner \etal  2005b;  Wang  \etal  2005).  Since  the
orientation   of  the  halo   itself  is   largely  governed   by  the
directionality  of its  mass accretion  (e.g., van  Haarlem \&  van de
Weygaert 1993; Tormen 1997),  this naturally explains the alignment of
the subhaloes with the major axis of the parent halo.  It is important
to distinguish  here between simple angular anisotropy  of the subhalo
distribution, and  a true  angular bias.  Even  in the absence  of any
spatial bias, any non-sphericity of  dark matter haloes will result in
a projected,  angular anisotropy  of the subhalo  distribution, unless
the halo is seen along its  symmetry axis.  Both Wang \etal (2005) and
Agustsson   \&   Brainerd  (2005a)   have   shown   that  the   halo's
non-sphericity  is   the  main   cause  of  the   theoretical  angular
anisotropy, but  that there is  a weak indication for  some additional
angular bias with a preferred alignment along the halo's major axis.

If the orientations of central galaxies are somehow aligned with their
dark  matter haloes, this  anisotropy should  result in  an observable
correlation  between the  distribution of  satellite galaxies  and the
orientation  of their central  galaxy.  Numerical  simulations suggest
that the angular  momenta of dark matter haloes  are typically aligned
with  their  minor  axes  (e.g.,  Warren \etal  1992;  Dubinski  1992;
Porciani,  Dekel   \&  Hoffman   2002a;  Bailin  \&   Steinmetz  2005;
Faltenbacher \etal  2005; Wang \etal  2005).  If the  angular momentum
vector of the baryonic material is  well aligned with that of the dark
matter, one would  thus naively expect the spin  axes of disk galaxies
to  be aligned  with the  minor  axes of  their host  haloes, and  the
satellite  galaxies to  be  preferentially oriented  along the  disk's
major  axis. However,  detailed hydro-dynamical  simulations  reveal a
more complicated picture. First of all, even in the absence of cooling
the  spin axes  of the  baryons and  the dark  matter are  only poorly
aligned, with  a median misalignment angle of  about $\sim 20^{\circ}$
to  $  \sim 30^{\circ}$  (van  den Bosch  \etal  2002;  Chen, Jing  \&
Yoshikaw 2003; Sharma \& Steinmetz 2005).  Furthermore, in simulations
of disk galaxy formation that  include cooling, the orientation of the
disk spin axis is found to be virtually uncorrelated with the original
(i.e., in the absence of baryons) minor axis of the halo (Bailin \etal
2005).  The formation  of the disk modifies the  shape and orientation
of  the  inner  halo,  but   leaves  the  outer  halo  largely  intact
(Kazantzidis \etal  2004; Bailin \etal 2005).   Consequently, the disk
spin axis is  well aligned with the halo minor axis  in the inner halo
($r \lta  0.1 r_{\rm  vir}$), but is  basically uncorrelated  with the
minor  axis at  larger  halo-centric radii.   If  correct, this  would
predict basically no alignment  between the orientation of the central
disk galaxy and the distribution  of its satellites (most of which lie
at relatively large halo-centric radii).

The  observational search  for  a possible  alignment between  central
galaxies  and their  satellites  has  a long  and  confusing history.  
Holmberg (1969) studied the  distribution of satellite galaxies around
isolated disk galaxies, and found them to lie preferentially along the
minor  axis of  disk  galaxies.  Holmberg's  study  was restricted  to
projected  satellite-central   distances  of  $r_p  \lta   50  \kpc$.  
Subsequent  studies, however,  were unable  to confirm  this so-called
``Holmberg-effect'' (Hawley \& Peebles 1975; Sharp, Lin \& White 1979;
MacGillivray \etal 1982).  Zaritsky  \etal (1997b) were also unable to
detect any  significant alignment  for $r_p \lta  200 \kpc$,  but they
{\it did} detect  a preferred minor axis alignment  for separations in
the range $300  \kpc \lta r_p \lta 500 \kpc$.   Note that this implies
an alignment  on scales  larger than the  typical virial radii  of the
haloes hosting these isolated disk galaxies.  This large-scale ($r_p <
500 \kpc$)  alignment has recently  been confirmed by Sales  \& Lambas
(2004), using  data from the two-Degree Field  Galaxy Redshift Survey,
but  only  for  host-satellite  pairs with  a  line-of-sight  velocity
difference of  $\vert \Delta v \vert  < 160 \kms$.   Our own Milky-Way
also reveals a Holmberg effect, in that the 11 innermost MW satellites
(with  MW  distances  $\lta   250  \kpc$)  show  a  pronounced  planar
distribution  oriented   close  to   perpendicular  to  the   MW  disk
(Lynden-Bell  1982;  Majewski 1994;  Kroupa,  Theis  \&  Boily 2005).  
Completely  opposite  to  all  theses  results,  Brainerd  (2005)  and
Agustsson \& Brainerd (2005) recently found that, in the Sloan Digital
Sky Survey  (SDSS), the distribution  of satellite galaxies  with $r_p
\lta 100  \kpc$ is strongly aligned  with the {\it major}  axis of the
disk host galaxy.  As a summary, we list in Table 1. the main attempts
in searching for the alignment signal between the central galaxies and
their satellites.

Clearly, this  lack of agreement calls  for a more  in-depth study. In
this paper we investigate the alignment between satellite galaxies and
their host galaxies using data  from the SDSS.  Our approach, however,
differs  substantially  from  all  previous studies.   First  of  all,
previous studies  only focused  on relatively isolated  disk galaxies,
which  has the disadvantages  that it  drastically reduces  the sample
size,  and that  one  only selects  haloes  in relatively  low-density
environments. In addition, satellite  galaxies were always selected in
a  fixed  metric  aperture  centered  on the  host  galaxy.   For  low
luminosity  hosts, which  reside in  low mass  haloes, this  metric is
often much  larger than the expected  virial radius of  the host halo.
In  this paper  we study  the host-satellite  alignment using  a large
sample  of galaxy groups.   No isolation  criteria are  applied, which
allows  us  to  (i)  achieve  much  better  statistics,  and  (ii)  to
investigate how  the alignment strength depends  on various properties
of the  host halo,  the host galaxy,  and the satellite  galaxies.  In
addition, we only focus on satellites that are located within the host
halo's virial radius  with projected satellite-central distances $r_p<
r_{\rm vir}$ and  satellite-central line-of-sight velocity differences
$|\Delta v| < v_{\rm vir}$,  where $r_{\rm vir}$ and $v_{\rm vir}$ are
the  virial radius  and virial  velocity dispersion  of the  host dark
matter halo, respectively.  In other words, we select satellites using
a variable  aperture size that  is motivated by  the mass of  the host
halo (i.e., the galaxy group).   This has the important advantage that
we can clearly separate small-scale alignment ($r < r_{\rm vir}$) from
large-scale  alignment ($r  >  r_{\rm vir}$).   Depending  on how  the
shapes and  angular momenta  of dark matter  haloes are  oriented with
respect to  their surrounding large scale structure,  the alignment on
these  two different  scales may  well  be very  different (see  e.g.,
Barnes \&  Efstathiou 1987; Porciani \etal 2002a,b;  Navarro, Abadi \&
Steinmetz  2004;  Bailin \&  Steinmetz  2005;  Trujillo, Carretero  \&
Patiri 2005).

This  paper is  organized  as follows.   In Section~\ref{sec:data}  we
present the data and our methodology. Section~\ref{sec:res} presents a
careful analysis  of the alignment between the  orientation of central
galaxies  and  the  distribution  of  their  satellite  galaxies.   In
particular,  we  show  how  the  alignment  strength  depends  on  the
luminosity and  color of  the galaxies,  and on the  mass of  the dark
matter  haloes.  In   Section~\ref{sec:comp}  we  present  a  detailed
comparison with previous  studies. Finally, in Section~\ref{sec:concl}
we summarize our results, and discuss their implications.

\section{Data and analysis}
\label{sec:data}

\subsection{Galaxy groups}
\label{sec:groups}

In order to address  the possible alignment between satellite galaxies
and the  central galaxy of  their dark matter  parent halo we  use the
SDSS galaxy group catalogue  of Weinmann \etal (2005)\footnote{In this
paper,  we refer the  brightest member  in each  group as  the central
galaxy,  while  all  other  members  as  satellite  galaxies.}.   This
catalogue  was constructed  from the  New York  University Value-Added
Galaxy Catalogue (NYU-VAGC; Blanton \etal 2005)
\footnote{http://wassup.physics.nyu.edu/vagc/\#download}   using   the
halo-based  group finder  developed  by Yang  \etal (2005a,  hereafter
YMBJ).  The  NYU-VAGC is based on  the SDSS Data  Release 2 (Abazajian
\etal  2004), but with  an independent  set of  significantly improved
reductions. We only consider the  galaxies with redshifts in the range
$0.01 \leq z \leq 0.2$ and with a redshift completeness $c > 0.7$
\footnote{Because  of  the   survey  selection  effects  (e.g.,  fiber
collisions, etc.),  not all galaxies in the  photometric catalogue are
spectroscopically observed and  thus their spectroscopic redshifts are
measured.}, resulting  in a  sample of $184,425$  galaxies with  a sky
coverage of $\sim 1950 \, {\rm deg}^2$.

In brief,  the YMBJ  group finder works  as follows.   First potential
group  centers   are  identified  using   a  Friends-Of-Friends  (FOF)
algorithm or an isolation criterion.  Next, the total group luminosity
is estimated  which is converted into  an estimate for  the group mass
using  an assumed mass-to-light  ratio. From  this mass  estimate, the
radius and  velocity dispersion of the corresponding  dark matter halo
are estimated  using the virial equations,  which in turn  are used to
select group members in redshift  space. This method is iterated until
group memberships converge.  Detailed  tests with mock galaxy redshift
surveys  have shown  that this  group finder  recovers groups  with an
average completeness  of $\sim 90\%$  and with an  interloper fraction
that is  smaller than  $\sim 20\%$. The  resulting group  catalogue is
insensitive  to  the initial  assumption  regarding the  mass-to-light
ratios, and the group finder  is more successful than the conventional
FOF  method in  associating galaxies  according to  their  common dark
matter haloes (see YMBJ for details).
\begin{figure}
\centerline{\psfig{figure=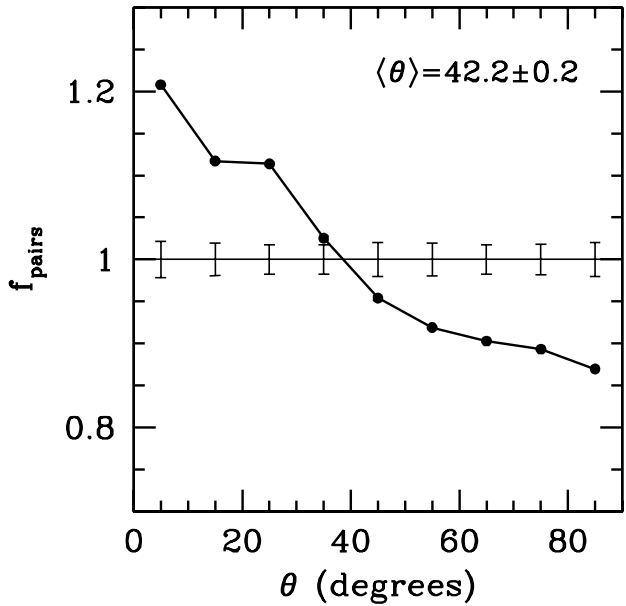,width=\hssize}}
\caption{The normalized probability distribution of the angle $\theta$ 
  between  the orientation  of the  major  axis of  the central  group
  galaxy  and the  direction of  each satellite  as measured  from the
  central galaxy. These results have been obtained from the SDSS group
  catalogue discussed in the text,  were we have excluded those groups
  for which  the projected ellipticity  of the central galaxy  is less
  than  $0.2$.    This  leaves  a  grand  total   of  $24,728$  unique
  central-satellite  pairs.   The   normalization  and  errorbars  are
  computed from  100 random  samples in which  we have  randomized the
  orientation  of all central  galaxies (see  text for  details). Note
  that $f_{\rm pairs}  > 1$ for $\theta <  35^{\circ}$ indicating that
  the  satellite  galaxies are  preferentially  distributed along  the
  major axis of  their central galaxy.  This is  also evident from the
  average value of $\theta$, and its error, which are indicated in the
  upper-right  corner. Note that  an isotropic  satellite distribution
  corresponds to $\langle \theta \rangle = 45^{\circ}$.}
\label{fig:orient}
\end{figure}

Following Yang  \etal (2005b)  we use the  group luminosity  to assign
masses  to  our  groups.   The  motivation behind  this  is  that  one
naturally expects the group  luminosity to be strongly correlated with
halo mass (albeit  with a certain amount of  scatter).  For each group
we determine the number density  of all groups brighter than the group
in consideration, using a common, empirically calibrated definition of
group  luminosity.  From  the halo  mass function  corresponding  to a
$\Lambda$CDM     concordance     cosmology    with     $\Omega_m=0.3$,
$\Omega_{\Lambda}=0.7$, $h=H_0/(100 \kmsmpc) = 0.7$ and $\sigma_8=0.9$
we then find the mass for  which the more massive haloes have the same
number density.   Although this  has the downside  that it  depends on
cosmology,  as shown  in  Weinmann \etal  (2005),  this method  yields
masses that are more accurate than those based on the more traditional
line-of-sight velocity  dispersion of the group  members (see Weinmann
\etal 2005). In addition, it  is straightforward to convert the masses
derived here to any other cosmology (see Yang \etal 2005b).

Applying our group  finder to the sample of  $184,425$ galaxies in the
NYU-VAGC described above yields  a group catalogue of $53,229$ systems
with  an estimated  mass. These  groups  contain a  total of  $92,315$
galaxies.  The majority of  the groups ($37,216$ systems) contain only
a single member, while there are $9220$ binary systems, $3073$ triplet
systems,  and $3720$  systems with  four members  or more\footnote{The
  group       catalogue      is       publicly       available      at
  http://www.astro.umass.edu/$^{\sim}$xhyang/Group.html}.    In   what
follows, we use this group  catalogue to examine the alignment between
the orientation of the central  galaxy, defined as the brightest group
member, and the distribution of  satellite galaxies. Note that we have
a total of $39,086$ unique  central-satellite pairs, which is an order
of magnitude larger than in any previous study.

\subsection{Methodology}
\label{sec:method}

In order to quantify the  distribution of satellite galaxies in groups
relative  to the  orientations  of their  central  galaxies we  follow
Brainerd  (2005) and compute  the distribution  function, $P(\theta)$,
where $\theta$ is  the angle on the sky between the  major axis of the
25 magnitudes  per square  arcsecond isophote in  the $r$-band  of the
central group galaxy and the  direction of a satellite relative to the
central    galaxy.     We    restrict    $\theta$   to    the    range
$[0^\circ,90^\circ]$, where $\theta=0^\circ$ ($90^\circ$) implies that
the satellite lies along the major (minor) axis of the central galaxy.
The  orientation of  the  central  galaxy is  based  on the  isophotal
position angle  in the $r$-band, as  given in the SDSS  Data Release 2
(Abazajian \etal 2004).
\begin{figure*}
\centerline{\psfig{figure=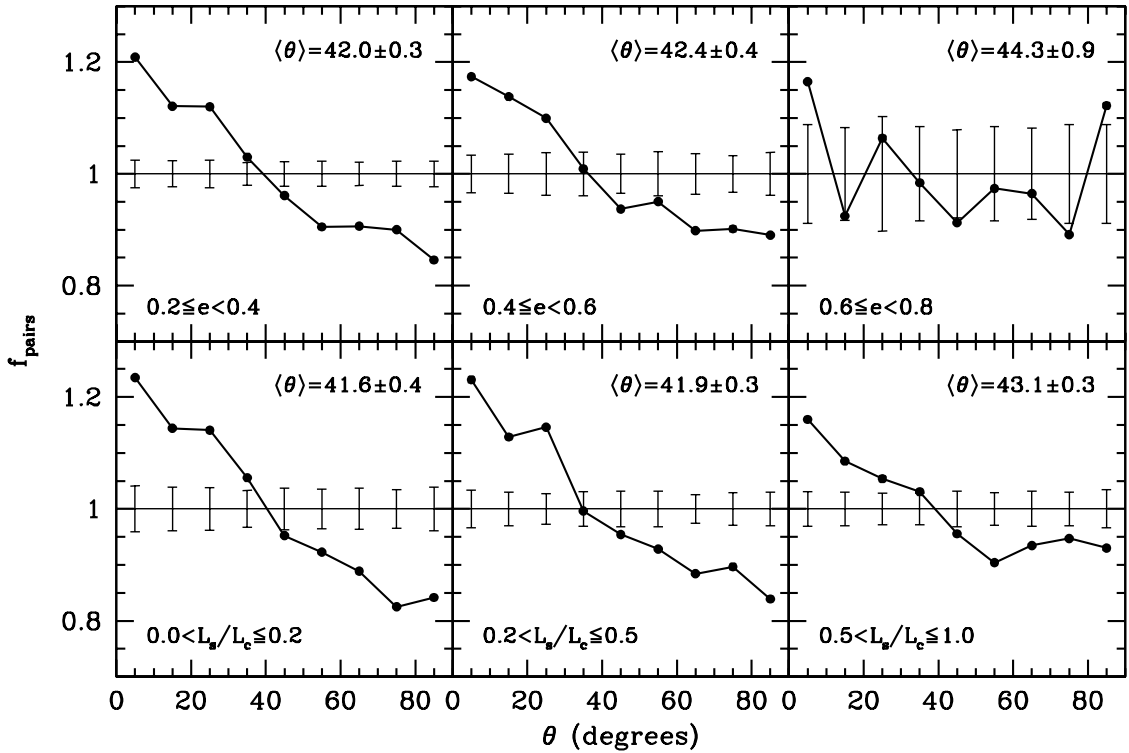,width=0.85\hdsize}}
\caption{The same as Fig.~\ref{fig:orient} but for different
  subsamples of  central and satellite galaxies. In  the upper panels,
  we  show  $f_{\rm  pairs}(\theta)$   for  groups  with  a  different
  ellipticity, $e$,  of the central  galaxy, as indicated.   Note that
  groups with a strongly elongated central galaxy ($0.6 \leq e < 0.8$)
  are   consistent  with   a  perfectly   isotropic   distribution  of
  satellites.    As   we   argue    in   the   text,   and   show   in
  Fig~\ref{fig:color}, this  owes to the fact  that strongly elongated
  systems  are mainly  blue, late  type disk  galaxies, which  show no
  significant   alignment.   The   lower  panels   show   how  $f_{\rm
    pairs}(\theta)$  depends  on  the  luminosities of  the  satellite
  galaxies,  $L_s$, expressed  in  units of  the  luminosity of  their
  central  galaxy, $L_c$.  There is  a clear  indication  that fainter
  satellites are more strongly aligned.}
\label{fig:ell_lum}
\end{figure*}

As  mentioned  above,  the  analysis  of Brainerd  (2005)  focused  on
relatively  isolated  systems with  late-type  central galaxies.   Our
analysis is  different in  that we consider  galaxy groups  of various
properties.  In practice,  we start with a group  sample and count the
total number of central-satellite  pairs, $N(\theta)$, for a number of
bins in  $\theta$. Next  we construct 100  random samples in  which we
randomize  the orientation  of all  central galaxies,  and  we compute
$\langle N_R(\theta) \rangle$, the average number of central-satellite
pairs as function of $\theta$.  Note that this ensures that the random
samples have exactly the same selection effects as the real sample, so
that any significant  difference between $N(\theta)$ and $N_R(\theta)$
reflects a  genuine alignment between  the orientation of  the central
galaxies and the distribution of satellite galaxies.

To  quantify the  strength of  any  possible alignment  we define  the
normalized pair count,
\begin{equation}
f_{\rm pairs}(\theta) = N(\theta) / \langle N_R(\theta) \rangle \,,
\end{equation}
Note that  $f_{\rm pairs}(\theta)=1$ in the absence  of any alignment.
We  use  $\sigma_R(\theta)   /  \langle  N_R(\theta)  \rangle$,  where
$\sigma_R(\theta)$ is the standard deviation of $N_R(\theta)$ obtained
from  the  100 random  samples,  to  assess  the significance  of  the
deviation of  $f_{\rm pairs}(\theta)$ from unity. In  addition to this
normalized  pair count,  we also  compute the  average  angle $\langle
\theta  \rangle$. In  the  absence of  any  alignment $\langle  \theta
\rangle = 45^{\circ}$, however,  $\langle \theta \rangle = 45^{\circ}$
does  not mean  an  isotropic distribution.  The  significance of  any
alignment can  be expressed in  terms of $\langle \theta  \rangle$ and
$\sigma_{\theta}$,  the  variance  in  $\langle \theta  \rangle_R$  as
obtained from the 100 random samples.

Finally, since the accuracy with which $\theta$ can be measured scales
with the projected ellipticity of the central galaxy, we only consider
groups for which the ellipticity of  the central galaxy, $e \geq 0.2$. 
Here $e$ is defined as one minus the ratio between the minor and major
axes  of  the 25  magnitudes  per  square  arcsecond isophote  in  the
$r$-band  of  the  image  of  the  central  galaxy.  This  ellipticity
constraints brings the total  number of unique central-galaxy pairs to
$24,728$

\section{Results}
\label{sec:res}

Fig.~1 shows $f_{\rm pairs}(\theta)$ for  all groups in our SDSS group
catalogue with inferred halo masses of $M \geq 10^{12}\msunh$ and with
central galaxies that have $e>0.2$. Note the pronounced enhancement of
pairs  with  small  $\theta$,  implying that  satellite  galaxies  are
preferentially  distributed  along the  major  axes  of their  central
galaxies.  This  is also  evident from the  fact that  $\langle \theta
\rangle = 42.2^{\circ} \pm  0.2^{\circ}$, which deviates from the case
of  no alignment (i.e.,  $\langle \theta  \rangle =  45.0^{\circ}$) by
$14\sigma$!

Since the  accuracy of  the orientation angle  of a central  galaxy is
smaller for central galaxies that  appear rounder, the strength of the
alignment  may  be  diluted  due  to central  galaxies  with  a  small
ellipticity,  $e$.  In  order  to address  the  impact of  $e$ on  the
strength   of   the   alignment    signal,   the   upper   panels   of
Fig.~\ref{fig:ell_lum}  show $f_{\rm  pairs}(\theta)$ for  groups with
central galaxies with different ellipticities, as indicated. Note that
the  alignment strength  is weakest  for the  sample with  the highest
ellipticities  ($0.6 \leq  e <  0.8$).  This  is surprising  since one
would expect the orientation angle of these central galaxies to be the
most accurate. However, as  we will see in Section~\ref{sec:gax_prop},
the strength of the alignment is significantly weaker for systems with
blue, late-type central galaxies than for systems with red, early-type
central galaxies.  The  dependence on $e$ found here  is simply due to
the  fact that central  galaxies with  $e \geq  0.6$ are  dominated by
blue, late-type,  disk galaxies.   In what follows  we always  use all
groups with central galaxies with $e\ge 0.2$.

\subsection{Dependence on Galaxy Properties}
\label{sec:gax_prop}

Since our group  catalogue contains a large number  ($\sim 39,000$) of
central-satellite  pairs, it  allows  us to  study  how the  alignment
depends on various  properties of the central and  satellite galaxies. 
We  start by  examining  the  dependence on  the  luminosities of  the
satellite  galaxies.  The lower  panels of  Fig~\ref{fig:ell_lum} show
$f_{\rm pairs}(\theta)$ for a  number subsamples of satellite galaxies
that are selected based on  their luminosities, $L_s$, relative to the
luminosities of their central galaxies,  $L_c$.  There is a weak trend
that  fainter satellite galaxies  are more  strongly aligned  with the
orientation of their central  galaxy than brighter satellite galaxies. 

Next,  we  consider the  dependence  on  the  color of  the  satellite
galaxies.  We separate galaxies into two subsamples according to their
$^{0.1}(g-r)$   colors,  which  corresponds   to  the   $(g-r)$  color
$k$-corrected to redshift $z=0.1$.  We call galaxies with $^{0.1}(g-r)
< 0.83$ `blue' and galaxies  with $^{0.1}(g-r) \geq 0.83$ `red'.  This
value of  $0.83$ roughly  corresponds to the  bimodality scale  in the
color-magnitude relation (see Weinmann \etal 2005).

\begin{figure}
\centerline{\psfig{figure=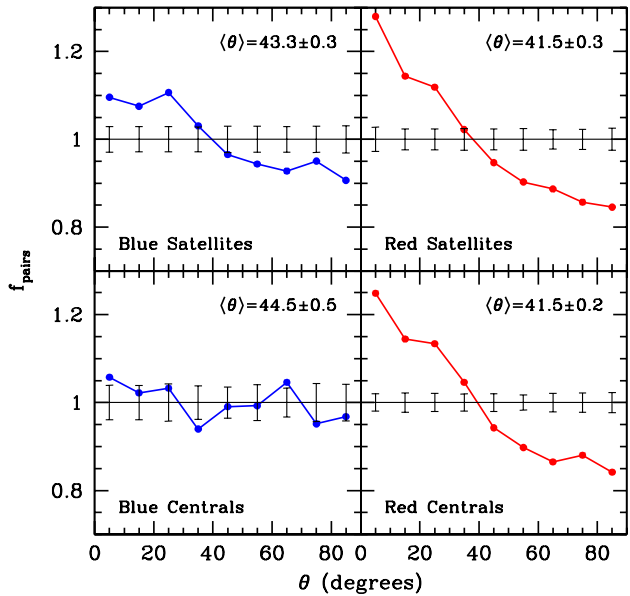,width=\hssize}}
\caption{The same as Fig.~\ref{fig:orient}, but for different
  subsamples  of hosts  and  satellites, selected  according to  their
  $^{0.1}(g-r)$ color. See text for discussion.}
\label{fig:color}
\end{figure}

The upper panels  of Fig.~\ref{fig:color} show $f_{\rm pairs}(\theta)$
for blue and red {\it satellite} galaxies, while the lower panels show
$f_{\rm  pairs}(\theta)$ for groups  with blue  and red  {\it central}
galaxies. Note  that there  is a remarkably  strong dependence  on the
color of  both the  central galaxies and  the satellite  galaxies.  In
particular, satellite  galaxies in groups with a  blue, central galaxy
are consistent  with a perfectly  isotropic distribution; there  is no
sign  of   any  significant  alignment  ($\langle   \theta  \rangle  =
44.5^{\circ}  \pm 0.5^{\circ}$). On  the contrary,  groups with  a red
central  galaxy  show a  very  pronounced,  major-axis alignment  with
$\langle \theta \rangle = 41.5^{\circ} \pm 0.2^{\circ}$.  In addition,
red satellites show a significantly stronger major axis alignment than
blue satellites.

\begin{figure}
\centerline{\psfig{figure=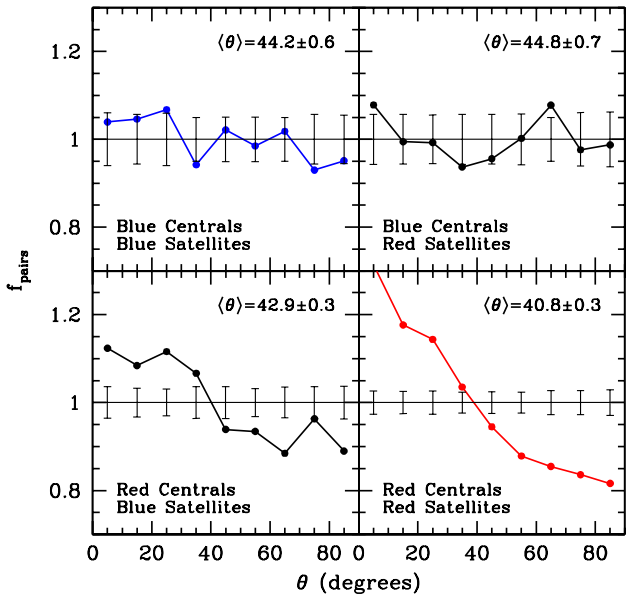,width=\hssize}}
\caption{Same as Fig~\ref{fig:color}, except that here we split
  the sample according to the colors of {\it both} the central and the
  satellite galaxies, as indicated.}
\label{fig:both_color}
\end{figure}

As shown  in Weinmann \etal (2005),  haloes with a  central red galaxy
have a significantly larger fraction  of red satellites than a halo of
the  same  mass, but  with  a  blue  central galaxy.   This  so-called
`galactic conformity' implies that the  upper and lower panels are not
independent.   In Fig.~\ref{fig:both_color}  we therefore  examine how
$f_{\rm  pairs}(\theta)$  depends on  the  colors  of  {\it both}  the
central galaxy  and the  satellites.  As can  be seen, systems  with a
blue central galaxy show  no significant alignment, neither with their
blue satellites  nor with  their red satellites.   Systems with  a red
central galaxy,  however, show a  very pronounced alignment,  which is
significantly  stronger  for  red  satellites  than  it  is  for  blue
satellites.   Since  redder colors  typically  indicate older  stellar
populations,  these  results  suggest  that  a  significant  alignment
between the  orientation of central  galaxies and the  distribution of
their satellite galaxies  only exists in haloes with  a relatively old
stellar population. Clearly, such  a correlation between the alignment
strength  and  the  age  of  the stellar  population  must  hold  some
interesting  clues regarding galaxy  formation. We  return to  this in
Section~\ref{sec:concl}.

\subsection{Halo Mass Dependence} 
\label{sec:mass}

\begin{figure*}
\centerline{\psfig{figure=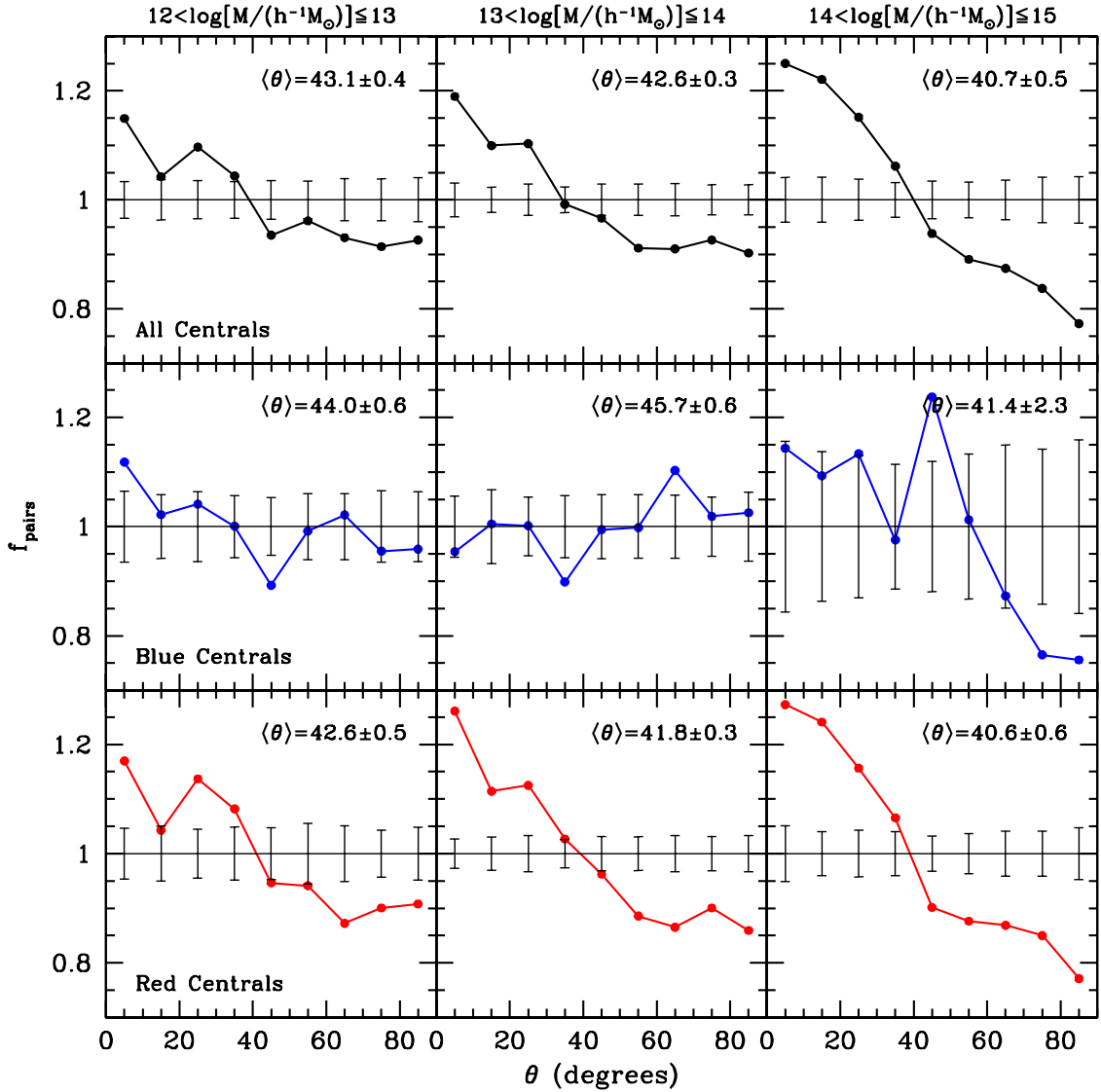,width=0.85\hdsize}}
\caption{The same as Fig~\ref{fig:orient}, but for
  satellite-central galaxy  pairs in haloes of different  mass bins as
  indicated at the top of the  panels. In the upper panels we show the
  results using all groups with $e \geq 0.2$, independent of the color
  of  the  central  galaxies. Note  that  there  is  a weak  trend  of
  increasing alignment  strength with increasing halo  mass. Panels in
  the  middle and lower  rows correspond  to haloes  in the  same mass
  ranges but  with only blue  and red central galaxies,  respectively. 
  Note  that haloes with  blue, central  galaxies show  no significant
  alignment of  their satellite  distribution with the  orientation of
  the central  galaxy, independent of  halo mass.  Haloes with  a red,
  central  galaxy  on the  other  hand,  always  reveal a  major  axis
  alignment, with a strength that increases with halo mass.}
\label{fig:halo_mass}
\end{figure*}

It  is interesting  to  examine whether  the  alignment strength  also
depends on halo mass.  Since  our group catalogue covers a large range
in  halo masses, we  can address  this question  in some  detail.  The
upper panels in Fig~\ref{fig:halo_mass} show the results for groups in
3 mass bins,  as indicated.  There is a  clear mass-dependence, in the
sense that the alignment is stronger for more massive groups.

Since more  massive haloes contain  a larger fraction of  red galaxies
(e.g., Weinmann \etal  2005), and given that red  galaxies show a much
more pronounced alignment than blue galaxies, this mass dependence may
simply reflect the color dependence shown in the previous section. To
test   this,   the  panels   in   the   middle   and  lower   row   of
Fig~\ref{fig:halo_mass} show $f_{\rm pairs}(\theta)$ for groups in the
same three mass bins, but  considering groups with blue or red central
galaxies  separately.  This  shows  that haloes  with  a blue  central
galaxies  show no significant  alignment, independent  of their  mass. 
Haloes  with a  red central  galaxy,  however, do  show a  significant
alignment, with a  strength that increases with increasing  halo mass. 
Thus, there is a genuine mass dependence, but only for haloes with red
central galaxies.

\subsection{Radial Dependence}
\label{sec:rad}

\begin{figure*}
\centerline{\psfig{figure=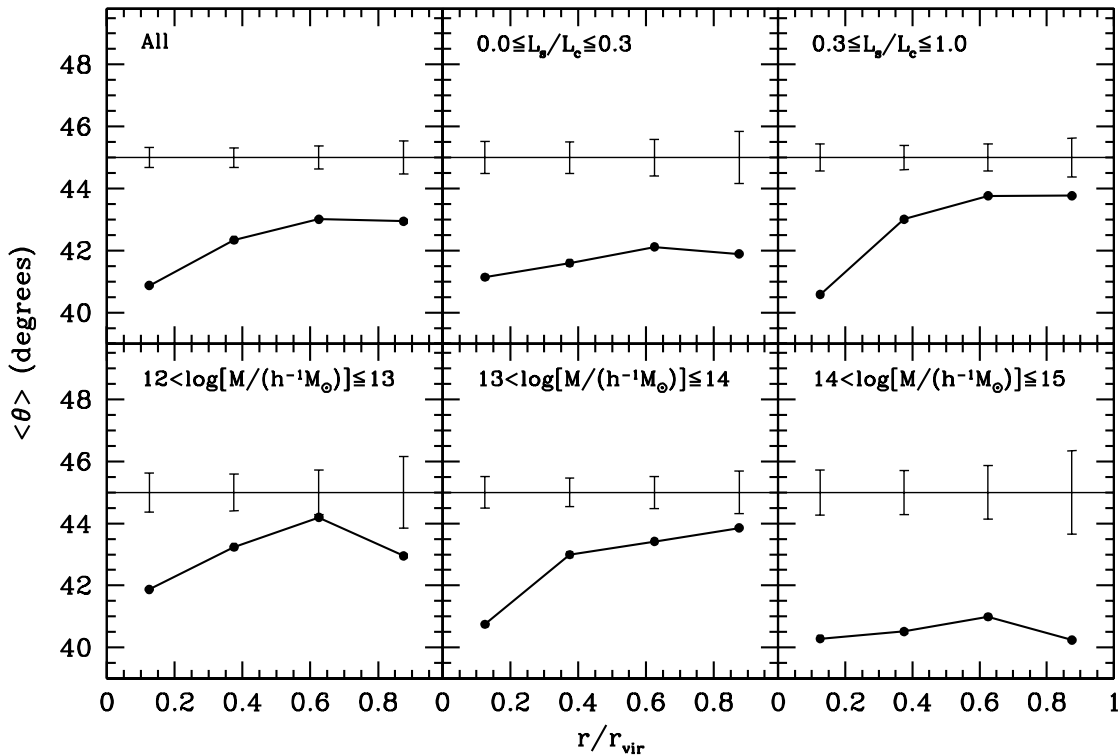,width=0.85\hdsize}}
\caption{The average angle $\theta$ as function of the projected
  radius between  the satellite galaxy  and the central  group galaxy,
  $r$, expressed in units of the group's virial radius, $r_{\rm vir}$.
  The upper left-hand  panel shows the result for  all groups in which
  the central galaxy has an ellipticity $e \geq 0.2$. The upper panels
  in the middle and to the right correspond to central-satellite pairs
  for $L_s/L_c$  falls in the  range indicated. The lower  panels show
  the results for  three subsamples selected according to  the mass of
  the groups, again as  indicated. The thin, horizontal line indicates
  $\langle  \theta  \rangle =  45^{\circ}$,  which  corresponds to  an
  isotropic distribution.  The errorbars  are computed from 100 random
  samples  in  which we  randomized  the  orientation  of all  central
  galaxies. Overall, the major  axis alignment strength is stronger at
  smaller projected radii.}
\label{fig:rad}
\end{figure*}

Finally, we examine whether the alignment depends on the group-centric
distance of  satellite galaxies.   For each central-satellite  pair we
compute the projected separation, $r$,  in units of the virial radius,
$r_{\rm  vir}$, of the  corresponding halo.   Fig.~\ref{fig:rad} plots
$\langle \theta \rangle$ as function of $r/r_{\rm vir}$.  An isotropic
satellite   distribution   will  have   $\langle   \theta  \rangle   =
45^{\circ}$, indicated  by the  horizontal line, while  values smaller
(larger)  than $45^{\circ}$  indicate a  preferred alignment  with the
major (minor) axis of the central galaxy. As before, the errorbars are
obtained from the 100 randomizations  of the orientations of the major
axes of  the central galaxies.   The upper, left-hand panel  shows the
results for all groups with central galaxies with $e \geq 0.2$.  There
is  a clear  radial trend,  in that  satellites at  smaller, projected
distances from their central galaxy are more strongly aligned with its
major  axis.   If we  split  the  sample  according to  the  satellite
luminosity,  $L_s$, relative  to that  of the  central  galaxy, $L_c$,
there is  a pronounced  difference: fainter satellites  do not  show a
significant radial  dependence, while satellites with $L_s  > 0.3 L_c$
show a very strong radial trend.

The lower  panels of Fig.~\ref{fig:rad} show  $\langle \theta \rangle$
as function of $r/r_{\rm vir}$  for three different bins in halo mass,
as  indicated. Haloes  with $M  \lta  10^{14} h^{-1}  \Msun$ reveal  a
significant  trend of  decreasing  $\langle \theta  \rangle$ (i.e.,  a
stronger major-axis alignment) with decreasing radius. In more massive
haloes,  however, there is  no significant  radial trend.  Instead, in
these  haloes the  major axis  alignment  is extremely  strong at  all
projected radii.

\begin{figure}
\centerline{\psfig{figure=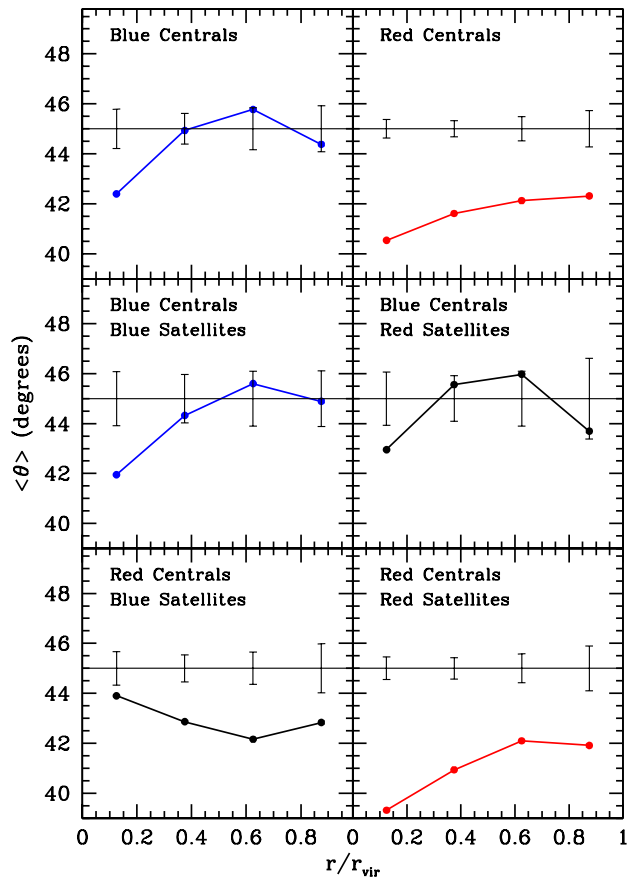,width=\hssize}}
\caption{Same as Fig.~\ref{fig:rad} except that we show the results
  for  different subsamples  selected according  to  the $^{0.1}(g-r)$
  colors of the central and satellite galaxies, as indicated. See text
  for a detailed discussion.}
\label{fig:radcol}
\end{figure}

Fig.~\ref{fig:radcol}  shows   how  $\langle  \theta  \rangle(r/r_{\rm
  vir})$ depends  on the  colors of the  satellites and  their central
galaxies.  Systems with a blue  central galaxy only show a weak ($\sim
3 \sigma$) major axis alignment with satellites (both red and blue) at
$r  \leq 0.2r_{\rm  vir}$. The  distribution of  satellites  at larger
projected radii is perfectly consistent with isotropic. Systems with a
red, central  galaxy reveal a  weak, but significant, radial  trend of
decreasing  alignment strength  with increasing  radius. This  is most
pronounced  for the  red  satellites, while  the  blue satellites  are
consistent (within  the errors) with a constant  alignment strength at
all projected radii.

\section{Comparison with Previous Studies}
\label{sec:comp}

As  shown  above,  we  detect  a significant  alignment  of  satellite
galaxies with the major axis of  their central host galaxy. This is in
qualitative  agreement  with the  recent  studies  of Brainerd  (2005,
hereafter  B05)  and Agustsson  \&  Brainerd  (2005b),  but in  strong
disagreement  with Holmberg (1969),  Zaritsky \etal  (1997b, hereafter
ZSFW97) and Sales \& Lambas (2004, hereafter SL04).

First of  all, given that many  studies have been  unable to reproduce
the  results of  Holmberg  (1969), and  given  that he  used a  sample
consisting  of  only  58  hosts  and 218  satellites,  we  argue  that
Holmberg's results  are probably an  unfortunate outcome of  the small
sample size.   Secondly, our results are  not necessarily inconsistent
with  those of  ZSFW97, who  only  detected a  significant minor  axis
alignment at relatively  large projected radii ($300 \kpc  \leq r \leq
500 \kpc$).   This is larger  than the typical virial  radius, $r_{\rm
  vir}$, expected for the isolated  disk galaxies used in their study. 
Since we have  only focused on the alignment at  scales $r \leq r_{\rm
  vir}$, our  results and theirs  are not mutually exclusive.   For $r
\leq 200  \kpc$ ZSFW97 did not  find any indication  for a significant
satellite  alignment.   Since  they  only  focused  on  isolated  disk
galaxies,  this is  in agreement  with the  isotropic  distribution of
satellites in systems with a blue central galaxy presented here.  Note
that ZSFW97 only had a  sample consisting of 115 satellites (around 69
host galaxies).

The  results of  SL04,  based on  the  2dFGRS, are  more difficult  to
explain  in   light  of  our  findings.   In   particular,  SL04  also
investigated  how   their  alignment  strength   correlates  with  the
properties of  the central galaxies.   In agreement with  our results,
they find  that the satellite  distributions around blue  centrals are
consistent with being isotropic,  while satellites around red centrals
show a strong  alignment effect.  However, contrary to  the major axis
alignment found here, SL04 detected a {\it minor} axis alignment.  The
fact that the same trends are detected, but in the opposite direction,
is suggestive of an error in  the computation of $\theta$, and we have
performed  a  number  of   tests  to  investigate  this  possibility.  
Unfortunately, the  major axis position angles of  the 2dFGRS galaxies
are  defined   as  ``measured  in  degrees  clockwise   from  East  to
West''\footnote{http://www.mso.anu.edu.au/2dFGRS/}.   This description
is  ambiguous  as it  is  unclear  whether  an angle  of  $45^{\circ}$
corresponds to  Northeast (as  would be the  case if  the astronomical
convention is used) or to Southeast.  To test this, we cross-correlate
the 2dFGRS  with our SDSS  sample, and compare the  orientation angles
provided  by  both catalogues.   This  comparison  indicates that  the
2dFGRS orientation angles are measured  from East through South. As it
turns  out, SL04 interpreted  the orientation  angles as  running from
East  through  North  (Laura  Sales  and Diego  Lambas,  {\it  private
  communication}).   Consequently,   what  they  call   a  minor  axis
alignment is in  fact a major axis alignment.  In retrospect, the SL04
results  are thus  in  qualitative  agreement with  B05  and with  the
results presented here. This has  been confirmed by tests performed by
the   authors   (Laura   Sales   and  Diego   Lambas,   {\it   private
  communication}).

Finally, it is worth pointing  out that our qualitative agreement with
the study  of B05, who also  used SDSS data, is  not entirely trivial.
While B05 focused on relatively isolated central galaxies, we consider
all galaxy systems,  from poor groups to rich  clusters, selected with
our group finder.   Furthermore, B05 only select systems  in which all
satellites are significantly fainter  than the central galaxy (similar
to  ZSFW97  and  SL04).   In  fact,  the  selection  criteria  are  so
restrictive, that  although B05 starts  out with a larger  SDSS sample
than used here  (SDSS Data Release 3 versus SDSS  Data Release 2), she
ends up with  less than 3300 satellites, almost  an order of magnitude
fewer  than in  our case.   Given  these dramatic  differences, it  is
therefore not obvious that the results have to be compatible.  We have
verified, however, that if we  select, from our group catalogue, those
groups that obey the selection criteria of B05, we obtain results that
are  of slightly  stronger alignment  signal than  those of  B05. This
discrepancy may partly due to  the fact that the color distribution of
the remaining central galaxies are  very similar to the original ones,
unlike  those in  B05 which  are mostly  spiral galaxies  and,  in our
investigations, have smaller alignment signal.

To  summarize, B05 and  SL04 both  have obtained  results that  are in
qualitative  agreement  with the  results  presented here:  Satellites
around  red  hosts are  aligned  with  the  host's major  axis,  while
satellites  around blue  hosts have  an angular  distribution  that is
consistent with  isotropic. As we  have argued, these results  are not
necessarily inconsistent with those of  ZSFW97. The only study that is
in clear  disagreement with  our results is  that of Holmberg  (1969). 
However, given  his relatively small sample size,  this discrepancy is
not very  significant.  Nevertheless,  one important exception  to the
general picture obtained here exists, namely our own MW.  As discussed
in  Section~\ref{sec:intro},  the MW  satellites  reveal a  pronounced
planar  distribution that is  oriented close  to perpendicular  to the
disk. However, as we argue below, this is not necessarily inconsistent
with our results.

\section{Discussion \& Conclusions}
\label{sec:concl}

Using  galaxy groups  selected from  the  SDSS, we  have examined  the
alignment between  the orientation of  the central galaxy  (defined as
the brightest  group member) and  the distribution of its  satellites. 
Overall we  find an excess of  satellites along the major  axis, and a
deficiency   along  the   minor   axis,  compared   to  an   isotropic
distribution.   The  alignment strength  in  our  sample is  strongest
between red central galaxies and  red satellites. On the contrary, the
satellite  distribution  in systems  with  a  blue  central galaxy  is
perfectly consistent  with isotropic. We also find  that the alignment
strength is stronger  in more massive haloes and  at smaller projected
radii from the central galaxy. In addition, there is a weak indication
that  fainter (relative  to the  central galaxy)  satellites  are more
strongly aligned.

Two conditions  must be  satisfied in order  to produce  the alignment
observed here. First of all, the distribution of satellite galaxies in
groups must  be aspherical, and  secondly, the orientation  of central
galaxies must be aligned with  the distribution of satellite galaxies. 
Cosmological  $N$-body  simulations of  CDM  models have  demonstrated
clearly that CDM haloes  are not spherical. The typical minor-to-major
axis ratio  is $\sim  0.6$, with a  relatively large dispersion  (e.g. 
Bullock 2002;  Jing \&  Suto 2002). This  is also supported  by recent
weak-lensing data  (Hoekstra, Yee \&  Gladders 2004). As we  argued in
Section~\ref{sec:intro},   simulations   suggest   that  the   angular
distribution  of dark  matter subhaloes,  which are  expected  to host
satellite galaxies, is  in reasonable agreement with that  of the dark
matter.   Therefore, it  seems  reasonable to  expect  that the  first
condition is  fulfilled. Explaining  the anisotropy as  reflecting the
non-sphericity of the  dark matter haloes is also  consistent with our
finding that  the alignment strength  increases with halo  mass; after
all,  numerical simulations have  shown that  more massive  haloes are
less spherical  (Warren \etal 1992;  Bullock 2002; Jing \&  Suto 2002;
Bailin \& Steinmetz 2005; Kasun \& Evrard 2005).

In order for the second condition to be fulfilled as well, the central
galaxy must  be somehow  aligned with the  principal axes of  the mass
distribution of  its host halo.   Here it is important  to distinguish
between disk galaxies, whose  orientation is governed by their angular
momentum vector, and spheroidal galaxies, whose orientation is somehow
related   to  its   formation   history  (typically   thought  to   be
merger-driven).

In the  standard picture of  disk formation (e.g., Fall  \& Efstathiou
1980; Mo, Mao \& White 1998), one assumes that baryons and dark matter
have  identical distributions  of  specific angular  momentum (due  to
tidal  torques from  the  cosmological density  field),  and that  the
baryons conserve their specific  angular momentum when cooling to form
a centrifugally supported disk.  Since simulations have shown that the
spin axis of dark matter haloes  is well aligned with the halo's minor
axis, this simple  picture predicts that the disk  spin axis should be
parallel to the minor axis of the halo, and thus that the distribution
of  satellites  is  aligned   with  the  disk  major  axis.   Somewhat
surprisingly, disk galaxies (which are typically blue) are exactly the
subsample  that do  not seem  to reveal  a significant  alignment with
their satellites.

Detailed hydrodynamical simulations, however, have shown that the spin
axes of  the baryons and the  dark matter (in the  absence of cooling)
are only  poorly aligned (van den  Bosch \etal 2002;  Chen \etal 2003;
Sharma \& Steinmetz 2005). In  addition, if cooling is included in the
simulations, the resulting disks are  found to have spin axes that are
very poorly aligned  with the {\it original} (i.e.,  in the absence of
baryons) minor  axis of the  halo (Bailin \etal 2005).   This suggests
that the  (direction) of  the angular momentum  of the baryons  is not
well conserved  during the disk  formation process, and  thus predicts
that there is  little if any alignment between  the orientation of the
disk and that of its  satellite distribution. However, there may still
be a  clear alignment between  the distribution of the  satellites and
the principal  axes of  the dark matter  halo.  This picture  not only
explains the lack  of a significant alignment when  stacking many disk
galaxies, but also  the pronounced alignment found for  the MW system:
it only  requires that  the major axis  of the  MW halo happens  to be
oriented along  the spin axis  of the MW  disk.  However, in  a recent
merger-driven disk  formation theory of Robertson et  al.  (2005) that
the  disk  galaxies can  be  produced  through  high angular  momentum
accretion  of gas  rich progenitors,  the satellite  galaxies  will be
preferentially  aligned with  the disk  plane. Note  also that  the MW
satellites  galaxies   have  an  average  luminosity   ratio  that  is
significantly smaller than  our central-satellite systems. Given this,
for  the MW  case,  as discussed  in  Kroupa, Theis  \& Boily  (2005),
another   plausible  explanation   of  the   MW   (dwarf)  satellites'
distribution is  that, if most of  the dwarves are not  of dark matter
dominated, but stem  from one initial gas-rich parent  satellite on an
eccentric near polar orbit that  interacted with the young MW, forming
tidal arms semi-periodically as its orbit shrank, this signal shall be
naturally expected.

Perhaps somewhat surprisingly, we find the strongest alignment between
red central galaxies and red  satellites.  If the orientation of a red
(early-type)  central galaxy is  a reflection  of the  orbital angular
momentum  of the  progenitors that  merged during  its  formation, one
might naively  expect that this  orientation is also aligned  with the
major axis of the halo. After  all, the orientation of the halo itself
is largely governed by the directionality of its mass accretion (e.g.,
van Haarlem \& van de Weygaert  1993; Tormen 1997).  It is less clear,
however,  why  the  alignment  should  be so  much  stronger  for  red
satellites than for blue satellites. In the standard picture of galaxy
formation,  once a  satellite galaxy  has  been accreted  by a  bigger
system, its  gas reservoir  is stripped, resulting  in a  fairly quick
truncation of  star formation;  the galaxy will  become red.   In this
simple picture, one thus expects the color of a satellite galaxy to be
a reflection of the time since  it was accreted. If the orientation of
a halo,  and its population of  galaxies, changes as  function of time
with  respect  to  the  large  scale matter  distribution,  one  could
envision that those  satellites that were accreted at  around the same
time when the central galaxy  formed, show a more pronounced alignment
than those satellites that have  only recently been accreted. As shown
by Bailin  \& Steinmetz  (2004), most haloes  reveal some  slow figure
rotation, with  an amplitude  that can cause  a directional  change of
more than $90^{\circ}$ within a  Hubble time.  Figure rotation is thus
a  potential explanation  for  the satellite-color  dependence of  the
alignment strength.

Cosmological  $N$-body simulations  also show  that more  massive dark
matter haloes in general have  more elongated structures, and that the
iso-density  contours of  the dark  matter distributions  are strongly
aligned in  the inner part of a  halo (e.g.  Jing \etal  1995; Jing \&
Suto 2002; Bailin \& Steinmetz  2005).  Thus, if the central galaxy in
such  a halo  aligns  with the  inner part  of  the halo,  and if  the
distribution   of   satellite   galaxies   traces  the   dark   matter
distribution,  the major  axis of  the central  galaxy is  expected to
align with  the satellite distribution, and the  alignment is expected
to be  stronger on smaller  radii. This is  in agreement with  what we
found in this paper.

In  a  recent study,  Augustsson  \&  Brainerd  (2005a) used  the  GIF
simulations (e.g. Kauffmann 1999)  to interpret the observed alignment
signal  between the  host and  satellite  galaxies in  the context  of
structure  formation.   They  found  that the  alignment  between  the
satellites and  the host  halo major axis  is much stronger  than that
between  the host  and satellite  galaxies in  the  observations. This
would imply that the host galaxies are on some degree mis-aligned with
the host  halos. Meanwhile a  related interesting result  was recently
obtained by  Mandelbaum et al.  (2006a;b), who  used the galaxy-galaxy
weak  lensing  signals  in  the  SDSS observations  to  constrain  the
ellipticity  of dark  matter  haloes.  They  found  that, for  spirals
(lens),  the  ellipticity of  halo  and  light  is anti-aligned  on  a
1-2$\sigma$ level,  while for  ellipticals (lens), the  ellipticity of
halo  relative  to   light,  $f_h=e_{halo}/e_{light}$,  increase  with
luminosity.  Apparently,  these findings are helpful  to interpret the
overall central-satellite galaxy alignment signals we obtained in this
paper, especially the halo mass, luminsity and type dependences.

The discussion presented above shows  that the results obtained in the
present  paper are  in qualitative  agreement with  naive, theoretical
expectations.  In  order to  compare our results  to theory in  a more
quantitative manner, one has to  understand in more detail how central
galaxies in  different halos form, how the  formation processes affect
the orientations of  the central galaxies relative to  the haloes, and
how satellite galaxies trace  the mass distribution within dark matter
halos.   All these  issues require  detailed numerical  simulations of
galaxy formation  in the cosmic  density field. In addition,  to study
the color  dependence of the  alignment, some basic treatment  of star
formation will be necessary.  Semi-analytical techniques combined with
high-resolution numerical  simulations may be  particularly suited for
this purpose.  We  will return to these issues  in a forthcoming paper
(Kang et al., in preparation).

                                                                               
\section*{Acknowledgment}

We are  grateful to  Laura Sales  and Diego Lambas  for their  help in
exploring  the  origin  of  the alignment-discrepancy.  We  thank  the
referee, Pavel Kroupa, for  insightful comments that helped to improve
the  paper. Michael  Blanton is  acknowledged  for his  help with  the
NYU-VAGC,  and we  thank  Jeremy Bailin,  Mike  Irwin, Peder  Norberg,
Francisco  Prada, Andrew  Zentner, David  Hogg and  Morad  Masjedi for
useful discussions.  XY is supported  by the {\it One Hundred Talents}
project of the Chinese Academy  of Sciences and grants from NSFC.  YPJ
is  supported  by  the   grants  from  NSFC  (Nos.10125314,  10373012,
10533030) and   from   Shanghai  Key   Projects   in  Basic   research
(No. 04JC14079 and 05XD14019).
                                                                               

\end{document}